\documentclass[aps,prl,groupedaddress,twocolumn]{revtex4-1}
\pdfoutput=1

\usepackage{amsmath}
\usepackage{amssymb}
\usepackage{graphicx}
\usepackage{bm}
\usepackage{hyperref}
\hypersetup{%
breaklinks = {true},
citecolor = {blue},
colorlinks = {true},
linkcolor = {red},
urlcolor = {blue},
pdfauthor = {\textcopyright\ Johan Nilsson},
pdfcreator = {\LaTeX\ and \flqq hyperref\frqq},
pdffitwindow = {true},
pdfmenubar = {true},
pdfpagelayout = {SinglePage},
pdfstartview = {Fit},
pdftoolbar = {true},
plainpages = {false},
}

\usepackage{bbold}

\newcommand{\bbZ}{\mathbb{Z}}

\usepackage{color}
\usepackage{color}

\DeclareMathOperator{\per}{\text{per}}

\newcommand{\pca}{\mathcal P}

\begin{document}


\title{Physics-inspired derivations of some algorithms for computing the permanent}

\author{Johan Nilsson}
\affiliation{Department of Physics and Astronomy, Uppsala University,
P.O. Box 516, SE-75120, Uppsala, Sweden}
\email{johan.nilsson@physics.uu.se}

\date{June 21, 2017}

\begin{abstract}

We provide physics-inspired derivations of a number of algorithms for computing the permanent of a matrix.
In particular we formulate the computation of the permanent as a Grassmann integral that may be viewed as an interacting many-fermion problem.
Applying a discrete Hubbard-Stratonovich decoupling then gives approximation schemes that are equivalent to the familiar determinant Monte Carlo algorithm. This leads to elementary derivations of the well-known estimators of Godsil-Gutman and Karmarkar et al. Another straightfoward manipulation of the Grassmann integral, making use of gauge invariance, gives the efficient exact formula of Glynn.
In addition to these known results we also give some additional estimators and formulas that are natural in our formulation.

\end{abstract}

\maketitle



The permanent of an $n \times n$ matrix $A$ is defined by
\begin{equation}
\per A = \sum_\pca \prod_{i=1}^n A_{i \pca_i},
\end{equation}
where the sum is over all permutations $\pca$ of $1,2, \ldots, n$. The permanent is therefore somewhat similar to the determinant, but the signature of the permutation is absent in the formula. In physics the permanent appears for example when calculating the overlap of two bosonic many-body wave functions \cite{Negele_Orland}. It is widely appreciated that it is much more difficult to evaluate the permanent than the determinant. This was put on firm ground in the context of complexity theory when Valiant proved that even the computation of the permanent of a matrix with $0\,$-$1$ valued entries is in the class of $\#P$-complete problems \cite{valiant1979}. On a less sophisticated level one important reason for this complexity is that the permanent is not invariant under similarity transformations, whereas the determinant is.

Grassmann integrals are commonly used as a device in many-body quantum mechanics to study interacting fermions in the path integral formalism. They were introduced by Berezin \cite{BEREZIN_BOOK}, and are since long textbook material, see e.g. \cite{Negele_Orland}. The Gaussian Grassmann integral is of particular importance and results in a determinant
\begin{equation}
\int \bigl[d\xi^* d\xi \bigr]e^{-\sum_{i,j}\xi^*_i A_{ij} \xi_j} = \det A.
\label{eq:gaussian1}
\end{equation}
The permanent of a matrix can also be expressed as a Gaussian Grassmann integral
\begin{equation}
\per A = \int \bigl[ d\phi^* d\phi \bigr] e^{\sum_{i,j}\phi^*_i A_{ij} \phi_j},
\label{eq:gaussian2}
\end{equation}
over {\it even} Grassmann numbers that satisfy $\phi_i \phi_j = \phi_j \phi_i$ and $\phi_i^2=0$.
This simple formula is one of crucial importance for this letter. It has appeared before in the physics literature \cite{PALUMBO1994_1,PALUMBO1994_2}, but is apparently not well-known. The even Grassmann numbers have also been introduced in the applied mathematics  and computer science communities where they go under the name of ``zeons'', see e.g. \cite{Schott_Staples_BOOK}.
The proof of \eqref{eq:gaussian2} is easily obtained by direct expansion \cite{supplementary}.
Since different $\phi_i$ commute no sign is generated upon rearranging the $\phi$'s into canonical order after expanding the exponential.
The final result is therefore the permanent instead of the determinant.
For our purposes it is important to note that these Grassmann numbers can also be viewed as composite objects \cite{PALUMBO1994_1,PALUMBO1994_2,supplementary}, defined as products of two ordinary anti-commuting Grassmann numbers $\xi_i$ and $\eta_i$: $\phi_i = \xi_i \eta_i$.
We will also use that the integration measure for $\phi$ may be separated into a product of two independent ones over $\xi$ and $\eta$
\begin{equation}
\int \bigl[d\phi^* d\phi \bigr]
=
\int \bigl[d\xi^* d\xi \bigr] \int \bigl[d\eta^* d\eta \bigr]
.
\label{eq:measure}
\end{equation}
In the language of physics integrals of the type \eqref{eq:gaussian2} can describe an interacting fermion theory of doublons in the Hubbard model, and lattice models involving spin-1/2 objects and hard-core bosons \cite{nilssonunpublished2017}. In this case the matrix $A$ is of large dimension but has a specific block structure and is very sparse.

\subsection{Formulas from discrete Hubbard-Stratonovich transformations}

The Hubbard-Stratonovich (HS) transformation is another standard method used in many-body physics to reformulate an interacting theory in terms of a weighted sum of non-interacting ones \cite{Stratonovich_HS_1957,Hubbard_HS_1959}. This transformation is important both in analytical approaches as well as in the construction of the determinant quantum Monte Carlo formalism \cite{Blankenbecler1981}. For our Grassmann numbers we will first use a discrete variant of this transformation which is the simple identity
\begin{equation}
e^{\phi^* a \phi} = e^{\eta^* \xi^* a \xi \eta} = \frac{1}{2} \sum_{s = \pm 1} e^{s \sqrt{a} \xi^* \xi} e^{s \sqrt{a} \eta^* \eta} ,
\label{eq:HS_Z2}
\end{equation}
which is valid for any real or complex number $a$.
Similar discrete HS transformations can also be introduced at the operator level \cite{Hirsch_DiscreteHS1983}, but the Grassmann version is almost trivial in comparison.
Introducing a sign $s_{ij}$ for each of the $m$ non-vanishing matrix elements of $A_{ij}$ we may rewrite the exponential using
\begin{equation}
e^{\sum_{i,j}\phi^*_i A_{ij} \phi_j} = 
\frac{1}{2^m} \sum_{\{ S \}} 
e^{\sum_{ij} \xi^*_i G_{ij}(S) \xi_j}
e^{\sum_{ij} \eta^*_i G_{ij}(S) \eta_j},
\end{equation}
where the matrix elements of $G(S)$ are
\begin{equation}
G_{ij}(S) = \sqrt{A_{ij}} s_{ij},
\end{equation}
and $S$ is a shorthand for the particular configuration of all of the $s_{ij}$.
Using the behavior of the measure \eqref{eq:measure} we can now perform the Gaussian Grassmann integrals over $\xi$ and $\eta$ independently to get
\begin{equation}
\per A = \frac{1}{2^m} \sum_{\{S\}} \bigl( \det G(S) \bigr)^2.
\label{eq:GG}
\end{equation}
This is the unbiased Godsil-Gutman estimator for the permanent \cite{godsil1978matching}.
This estimator unfortunately has a very large variance in the worst case, and does therefore not work well for all matrices.

The Godsil-Gutman estimator has been generalized in many different ways to reduce its variance. In the physics language some such generalizations may be generated by using other HS transformations than the $\bbZ_2$ one of \eqref{eq:HS_Z2}. In particular we could instead use a $\bbZ_p$ decoupling
\begin{equation}
e^{\phi^* a \phi} = 
\frac{1}{p} \sum_{q=1}^{p} e^{\omega^q \sqrt{a} \xi^* \xi + \omega^{-q} \sqrt{a} \eta^* \eta},
\label{eq:HS_Zp}
\end{equation}
where $\omega = e^{2 \pi i /p}$ is a $p\,$-th root of unity. Note that \eqref{eq:HS_Z2} is a special case of this corresponding to $p=2$. Clearly this can again be done independently on each non-vanishing element of $A_{ij}$. Introducing $q_{ij}$ and $H(Q)$ in analogy with $s_{ij}$ and $G(S)$ above with
\begin{equation}
H_{ij}(Q) = \sqrt{A_{ij}} \omega^{q_{ij}},
\end{equation}
we may perform the Grassmann integrals with the result
\begin{equation}
\per A = \frac{1}{p^m} \sum_{\{Q\}} \bigl| \det H(Q) \bigr|^2,
\end{equation}
when $A$ has only positive semi-definite elements.
This includes (setting $p=3$) the KKLLL esimator of Karmarkar et al. \cite{Karmarkar}. The bounds on the variance of this estimator is much better than that of Godsil-Gutman. The variance is however still extremely large in the worst case \cite{Karmarkar}. On the other hand it has been proven that the KKLLL estimator is very efficient for sufficiently dense matrices \cite{Frieze1995}.

In the physics context it is well-known that it is possible to decouple interaction terms in different channels \cite{Negele_Orland}. This leads to additional formulas for the permanent.
So far we have only considered decouplings that preserves ``spin rotational symmetry'' around one axis, meaning that we have only made use of  bilinears of the types $\xi^*\xi$ and $\eta^*\eta$. In \eqref{eq:HS_Z2} we only  use the {\it density channel}, meaning that the bilinears are of the form $\xi^*\xi + \eta^*\eta$. It is also possible to do decouplings in the {\it spin channel}, the simplest of these involves $i(\xi^*\xi - \eta^*\eta)$, which do appear in  \eqref{eq:HS_Zp}. We may also decouple using the bilinears $i(\xi^*\eta + \eta^*\xi)$ and $(\xi^*\eta - \eta^*\xi)$, which breaks the spin rotational symmetry. These are all examples of time-reversal invariant decoupling schemes \cite{wu_zhang_positive}, which have the appealing property that the weights are always positive semi-definite when the elements of $A$ are. 
In fact it is also possible to use different forms of the decoupling on different matrix elements. A simple example demonstrates that using this freedom may be extremely fruitful. Consider the following matrix $A$ and one particular decoupled version $H(S)$
\begin{equation}
A =
\begin{pmatrix}
1 & 1 \\
1 &1
\end{pmatrix}
,
\qquad
H(S) =
\begin{pmatrix}
s_{11} & is_{12} \\
s_{21} & s_{22}
\end{pmatrix}
.
\label{eq:A22mat}
\end{equation}
The corresponding estimator has zero variance! Of course it is a very difficult problem to pick the ``best'' decoupling scheme for a given large matrix. In the physics context the choice of decoupling is supposed to be ``guided by the physics of the problem'' \cite{Negele_Orland}, something that seems quite difficult in the abstract setting of a generic matrix $A$.

It is also possible to perform a decoupling in the {\it pairing channel}, which is equivalent to a decoupling in $\phi^*$ and $\phi$. Let us use a general $\bbZ_p$ decoupling ($p \geq 2)$
\begin{equation}
e^{\phi^* a \phi} = \frac{1}{p} \sum_{q=1}^p e^{\omega^{q} \sqrt{a} \phi^*} e^{\omega^{-q} \sqrt{a} \phi} .
\label{eq:HS_Z2_pair}
\end{equation}
Implementing this on each non-vanishing term in $A$ as above and performing the Grassmann integral we obtain
\begin{multline}
\per A = \frac{1}{p^m} \sum_{\{Q\}} 
\\ \times
\prod_i \Bigl( \sum_{k_i} \sqrt{A_{k_i i}}\omega^{-q_{k_i i}} \Bigr) \Bigl( \sum_{j_i} \sqrt{A_{ij_i}}\omega^{q_{ij_i}} \Bigr) .
\end{multline}
Specializing to matrices with $0\,$-$1$ valued entries and taking $p=2$ we get
\begin{equation}
\per A = \frac{1}{2^m} \sum_{\{S\}} 
\prod_i \Bigl( \sum_{k_i} s_{k_i i} \Bigr) \Bigl( \sum_{j_i} s_{ij_i} \Bigr) .
\end{equation}
This formula treats rows and columns symmetrically and has a simple interpretation: all non-zero elements of $A$ are substituted with $\pm 1$ with equal probability, and the contribution from a configuration is the product of all row sums and columns sums.

\subsection{Efficient exact formulas}

Efficient exact formulas may also be easily obtained from manipulations of the Grassmann integral. Let us first note that because $\phi_i^{*2} = 0$ the Gaussian exponent may be expanded as
\begin{equation}
e^{\sum_{i,j}\phi^*_i A_{ij} \phi_j}
= \prod_i \Bigl(1 + \phi^*_i \sum_{j_i} A_{ij_i} \phi_{j_i} \Bigr).
\end{equation}
Sticking this into the formula for the permanent and performing the integrals over the $\phi_i^*$'s we get
\begin{equation}
\per A = \int \bigl[ d\phi \bigr] \prod_i \Bigl(\sum_{j_i} A_{ij_i} \phi_{j_i}\Bigr).
\end{equation}
Now let us attach a local $\bbZ_2$ gauge freedom to each $\phi_j$. We implement this by making a change of variables inside the integral, taking $\phi_j \rightarrow s_j \phi_j$ with $s_j = \pm 1$. The Jacobian generated by going to the new measure is simply $\prod_k s_k$. Since nothing can depend on the variable change, and hence the $s_i$'s, we may average over all possibilities
\begin{equation}
\per A = \frac{1}{2^n} \sum_{\{S\}} \Bigl(\prod_k s_k \Bigr) \int \bigl[ d\phi \bigr] \prod_i \Bigl(\sum_{j_i} A_{ij_i} s_{j_i}\phi_{j_i}\Bigr).
\end{equation}
In this formula the role of the Grassmann integral, i.e., to pick out the terms for which each $\phi_i$ appears exactly once, is superfluous since terms in the integrand without this property are anyway set to zero by the sum over $S$. Getting rid of the integral we therefore get
\begin{equation}
\per A = \frac{1}{2^n} \sum_{\{S\}} \Bigl(\prod_k s_k \Bigr) \prod_i \Bigl(\sum_{j_i} A_{ij_i} s_{j_i}\Bigr).
\label{eq:exactformula1}
\end{equation}
This is an exact formula that is  a sum of $2^n$ terms. A more efficient albeit less symmetric formula involving a sum of  $2^{n-1}$ terms may easily be obtained from this one by noting that each term is invariant upon inverting all signs. This implies that for each configuration with say $s_1 = -1$ there is another one with $s_1 = 1$ with equal weight. We may therefore write
\begin{eqnarray}
\per A = \frac{1}{2^{n-1}} {\sum_{\{S\}}}' \Bigl(\prod_k s_k \Bigr) \prod_i \Bigl(\sum_{j_i} A_{ij_i} s_{j_i}\Bigr),
\label{eq:exactformula2}
\end{eqnarray}
where the primed sum is over all sign configurations but with $s_1 = 1$ always. This is the formula of Glynn \cite{glynn2010}.
Glynn has also formulated this in terms of polarization identities \cite{glynn2013}, this makes clear the connection to the Ryser formula \cite{ryser1963}, which is of comparable efficiency.
It is interesting to note that in the physics language \eqref{eq:exactformula1} is a result of local $\bbZ_2$ gauge invariance, and the reduction to \eqref{eq:exactformula2} a consequence of global $\bbZ_2$ gauge invariance.

\subsection{Additional formulas}

The derivations above motivates us to consider a few generalizations that gives some additional formulas for the permanent.

It is clear that the formulas \eqref{eq:exactformula1} and \eqref{eq:exactformula2} may also be considered as unbiased estimators for the permanent when the signs are treated as random variables \cite{aaronson2014}. We now note that in the discrete HS case the variance was reduced in going from Godsil-Gutman to KKLLL, which may be viewed as changing the decoupling variable from $\bbZ_2$ to $\bbZ_p$ in our language. In the derivation of the exact formula above we may easily implement the same idea, i.e., we make the gauge transformation (variable change) $\phi_i \rightarrow \omega^{q_i} \phi_i$. This leads to the formula
\begin{equation}
\per A = \frac{1}{p^n} \sum_{\{Q\}} \Bigl(\prod_k \omega^{-q_k} \Bigr) \prod_i \Bigl(\sum_{j_i} A_{ij_i} \omega^{q_{j_i}}\Bigr).
\label{eq:exactformula3}
\end{equation}
Viewed as an exact formula this is obviously less efficient than \eqref{eq:exactformula1} since the number of terms is $p^n$ instead of $2^n$. From the point of view of an unbiased estimator the variance is however substantially reduced for certain matrices. 
A common example is the $n \times n$ block-diagonal matrix with the matrix $A$ in \eqref{eq:A22mat} repeated on the diagonal. For such a matrix the second moment is reduced by a factor of $(3/4)^{n/2}$ in going from \eqref{eq:exactformula1} to \eqref{eq:exactformula3} for all $p \geq 3$. This is similar to the reduction in going from Godsil-Gutman to KKLLL \cite{Karmarkar,chien_rasmussen_sinclair}.

It is also possible to consider various continuous decoupling schemes. Let us decompose our matrix as $A=LU\negthinspace P$ where $P$ is a permutation matrix and $L$ ($U$) are lower (upper) triangular matrices (such a decomposition is always possible). $P$ does not affect the permanent and we only need to consider the permanent of the matrix $LU$. Now we may perform a conventional continuous HS transformation on the exponent as follows
\begin{equation}
e^{\sum_{i,j,k}\phi_i^* L_{ij} U_{jk} \phi_k}
=
\int \bigl[d \varphi^* d \varphi\bigr] e^{\sum_{i,j} (\phi_i^* L_{ij}\varphi_j + \varphi_j^* U_{ji} \phi_i )},
\end{equation}
with a normalized Gaussian integration measure
\begin{equation}
\int \bigl[d \varphi^* d \varphi\bigr] 
=
\int \prod_j \frac{d \varphi_j^* d \varphi_j}{2\pi i} e^{-|\varphi_j|^2}.
\end{equation}
Sticking this into the formula for the permanent it is easy to perform the Grassmann integral with the result
\begin{equation}
\per A = \int \bigl[d \varphi^* d \varphi\bigr] 
\prod_i 
\Bigl(\sum_{k_i} \varphi^*_{k_i} U_{k_i i} \Bigr)
\Bigl(\sum_{j_i} L_{ij_i} \varphi_{j_i}\Bigr)
.
\label{eq:LUrep}
\end{equation}
This formula is a $2n$-dimensional integral representation of the permanent.
Similar formulas also works for other types of matrix decompositions. Consider for example a singular value decomposition $A = U \Sigma V^T$, with $U$ and $V$ orthogonal and $\Sigma$ diagonal. Let us denote the diagonal elements of $\Sigma$ by $\sigma_i$, then the same manipulations gives
\begin{multline}
\per A = \int \bigl[d \varphi^* d \varphi\bigr] 
 \\ \times
\prod_i \Bigl(\sum_{j_i} U_{ij_i} \sqrt{\sigma_{j_i}}\varphi_{j_i}\Bigr)
\Bigl(\sum_{k_i} V_{i k_i} \sqrt{\sigma_{k_i}}\varphi^*_{k_i} \Bigr).
\label{eq:SVDrep}
\end{multline}
These $2n$-dimensional integral representations involves an unbounded integration region. It is also possible to reformulate these on bounded integration regions, what is needed is a weight function that satisfy
\begin{equation}
 \int \bigl[d \varphi^* d \varphi\bigr] \varphi^{*i} \varphi^j   = \delta_{ij} j! \qquad \text{for } 0 \leq i,j \leq n .
\end{equation}
The integrals \eqref{eq:LUrep} and \eqref{eq:SVDrep} may be estimated using stochastic Monte Carlo methods. The last version is particularly efficient for matrices of low rank since the number of necessary $\varphi$'s is equal to the rank of the matrix.

Let us finally mention another possibility that may be obtained from a slight extension of our formalism.
Suppose that we represent $\phi = \mu \nu$, with $\mu$ and $\nu$ a pair of {\it even} Grassmann numbers. We may then decouple a term in the exponent as before
\begin{equation}
e^{\phi^* a \phi} = \frac{1}{2} \sum_{s = \pm 1} e^{s \sqrt{a} \mu^* \mu} e^{s \sqrt{a} \nu^* \nu} ,
\label{eq:HS_Z2_2}
\end{equation}
and the measure still factorizes in analogy with \eqref{eq:measure}.
Doing this for each of the $m$ non-vanishing elements of $A$ like we did to get to \eqref{eq:GG} and using \eqref{eq:gaussian2} we arrive at the formula
\begin{equation}
\per A = \frac{1}{2^m} \sum_{\{S\}} \bigl( \per G(S) \bigr)^2.
\end{equation}
This formula should be compared with \eqref{eq:GG} and can be used recursively to generate additional formulas.


\acknowledgements

\subsection{Acknowledgements}

Funding from the Knut and Alice Wallenberg Foundation and the Swedish research council Vetenskapsr{\aa}det is gratefully acknowledged.

\bibliography{permanent}

\end{document}